\pdfoutput=1
\documentclass[11pt,a4paper]{article}
\usepackage[margin=2.5cm]{geometry}
\usepackage{amsmath,amssymb}
\usepackage{graphicx,booktabs,array}
\usepackage[colorlinks=true,linkcolor=blue,citecolor=blue,urlcolor=blue]{hyperref}
\newtheorem{proposition}{Proposition}
\title{The forces at hand: social learning is permissive, not instructive,
for the evolution of phenotypic plasticity}
\author{Nam H. Le\\[2pt]
\small Department of Computer Science and Vermont Complex Systems Research Institute,\\
\small University of Vermont, Burlington, VT, USA\\
\small \texttt{nam.le@uvm.edu}\quad ORCID: 0000-0002-6862-9257}
\date{}

\begin{document}
\maketitle

\begin{abstract}
Evolution can favour a trait for two reasons: an instructive cause makes
selection favour the trait; a permissive one lets it persist once selection
no longer removes it. The distinction matters most for phenotypic plasticity
and its replacement by inherited behaviour, genetic assimilation, the Baldwin
effect. Social learning is credited with favouring plasticity on evidence
that cannot tell the two apart: how much plasticity remains at the end of a simulation. Here we show, analytically and computationally, which role it plays in a
minimal model of learning and evolution. Writing expected fitness as a
function of plasticity, we find that environmental change is the only source
of benefit from plasticity, and the social channel enters only its cost.
Simulation confirms this: in a static environment selection against
plasticity never reverses sign at any channel strength; when the environment
changes, it reverses at a threshold the channel lowers, and without the
channel the population collapses before reaching it. Social learning is
permissive, not instructive: it supplies the means to be plastic, never the
reason. For the debate over whether plasticity leads evolution, this is one
case settled by measurement: environmental change directs, socially inherited
information permits, and neither suffices alone.
\end{abstract}

\noindent\textbf{Keywords:} Baldwin effect; social learning; phenotypic
plasticity; genetic assimilation; genetic drift; Rogers' paradox

\section{Introduction}

Phenotypic plasticity, the capacity of one genotype to produce different
phenotypes in response to its environment, has moved to the centre of a
debate about the structure of evolutionary theory
\cite{westeberhard03,laland14,laland15,pfennig21}. Proponents of an extended
evolutionary synthesis argue that plasticity can lead adaptive evolution,
with environmentally induced phenotypes appearing first and genetic change
following, a sequence known as plasticity-led or plasticity-first evolution
\cite{levis16,levis20}; critics reply that such cases are already
accommodated by standard theory, in which selection on genetic variation
remains the cause of adaptation \cite{laland14}. Much of this disagreement is
about causation \cite{uller19}: whether plasticity helps to determine the
direction of evolution, or merely allows evolution to proceed in a direction
set by selection. Philosophers of biology call this the distinction between
instructive and permissive causes \cite{calcott,bourrat,gilbert}. Its
sharpest case is genetic assimilation, the replacement of a learned or
induced trait by an inherited one, and its best-known version, Baldwin's
organic selection \cite{baldwin,crispo07,scheiner17}: does learning steer
assimilation, or does it only keep the population alive while selection does
the steering?

The extended synthesis also widens what counts as inheritance, to include
information transmitted socially rather than genetically
\cite{danchin11,laland15}. Social learning is therefore where its two central
themes meet: it is a form of plasticity whose products are themselves
inherited. Because its effects on selection unfold over many generations, the
question has been studied mainly in computational models, in which some
aspects of an individual's phenotype are fixed by inheritance and others are
left open, to be settled by learning during its lifetime \cite{hn,nolfi94}.
Learning can then guide a population to adaptations that genetic search alone
would rarely find, and the amount of plasticity a population still carries at
the end of a simulation became the standard evidence in these models that learning
had been favoured by selection.

These models have since been extended to social learning, where an
individual's learning is biased by information taken from others, and the
extensions have repeatedly reported that this helps: Belew \cite{belew}
found that a cultural bias on what an individual learns raises the amount
of plasticity the population retains; Best \cite{best} reported
that copying a fitter individual's expressed behaviour can guide a
population to its target with no individual learning at all; Le and colleagues \cite{le19,le19gecco,le20evoapps} reported that a population
of social learners sustains more plasticity, and reaches higher average
fitness under a changing target, than individual learners alone; and in
evolutionary robotics, controllers copied between robots are credited with
a comparable adaptive advantage over evolution or individual learning on
their own \cite{heinerman,heinerman2,bredeche}. Read together, this body
of work has settled into a common claim: that adding a social channel to
this kind of model increases adaptation, and does so by increasing
plasticity.

That claim rests on two things that have not, until now, been checked.
First, on how the social channel is built. Most of the models above
implement social learning as a dedicated procedure, separate from the
individual learning it is compared against, which is precisely the
assumption that Heyes \cite{heyes12} argues against on independent
behavioural and comparative grounds: there is no dedicated social-learning
mechanism, only one learning mechanism supplied with input that is tuned,
developmentally or evolutionarily, towards other individuals. Section~\ref{sec:related}
reviews this literature in more detail and shows that the mismatch is the
rule rather than the exception. Second, even where that mismatch is set
aside, the claim rests on how much plasticity is retained, the same
diagnostic that cannot distinguish an instructive cause from a permissive
one. A population can end a run with more plasticity either because
selection has come to favour staying open, or because the social channel
has merely made staying open cheaper to carry; both produce the same
number, and no study in this tradition has measured which one occurred.

Here we show, analytically and computationally, which role the social
channel plays. We build a model in which social and individual information
enter one learning rule, as Heyes's criterion requires, and we measure
selection on plasticity directly, as the regression of an individual's
fitness on its own plasticity, rather than reading it off a final frequency.
Written as a function of plasticity, expected fitness has a single term that
rewards plasticity, and it belongs to the environment; the social channel
enters only the cost. The channel is therefore permissive, not instructive. A
sweep over channel strength and rate of environmental change confirms this,
and the same instruments resolve a twenty-year disagreement about whether
social learning alone can guide a population to its target (Appendix~\ref{app:bestle}).

The remainder of the paper is organised as follows. Section~\ref{sec:related}
reviews how social learning has been implemented and why it matters;
Section~\ref{sec:model} presents the model, the measurement and the analytic
criterion; Section~\ref{sec:results} reports the results; and
Section~\ref{sec:discussion} returns to the debate with which we began.

\section{Related work}
\label{sec:related}

Whether phenotypic plasticity can lead adaptive evolution is one of the
questions on which the proposed extension of the evolutionary synthesis
turns \cite{westeberhard03,laland14,laland15,pfennig21}. In the
plasticity-led account a new environment first induces a phenotype, the
induced phenotype exposes previously cryptic genetic variation, and selection
then refines and eventually fixes it, so that plasticity precedes and directs
the genetic change \cite{levis16,levis20}. The standard account replies that
nothing in this departs from selection acting on genetic variation: induction
may create the opportunity, but the direction of evolution is still set by
which variants selection favours \cite{laland14,futuyma17}. The oldest
version of the dispute concerns genetic assimilation and Baldwin's organic
selection \cite{baldwin,crispo07}, and even its modelling is unsettled:
Scheiner, Barfield and Holt \cite{scheiner17} found that assimilation as
originally proposed is too slow to be likely, while plasticity itself can
rescue populations from extinction under environmental change. What both
sides ask for is the same thing. Direct measurement of selection on
plasticity is among the criteria Levis and Pfennig \cite{levis16} set for a
plasticity-led case, and it is rarely available in nature; the computational models reviewed later in this section show that it has rarely been attempted even where it is cheap.

Underneath the dispute is a question about causal role. Developmental biology
distinguishes an instructive signal, which determines which of several
outcomes a competent tissue produces, from a permissive one, which the tissue
needs in order to produce an outcome already determined by something else
\cite{gilbert}. Woodward \cite{woodward10} gave the distinction a general
form: a cause is specific to an effect to the extent that interventions on it
map onto distinct values of the effect. Calcott \cite{calcott} showed that
instructive causes are the specific ones and permissive causes the
unspecific, and Bourrat \cite{bourrat} examined the limits of the contrast.
Uller and Laland \cite{uller19} argue that much of the extended-synthesis
debate is of this kind: both sides agree that development, inheritance and
environment are causally involved in evolution, and disagree about which of
them specify its direction. Put this way the question has a measurable form.
A cause whose intervention changes the sign of selection on a trait is
instructive with respect to that trait; one whose intervention changes only
the strength of selection, or whether selection can act at all, is
permissive. This paper builds a model in which both kinds of intervention can be made, on the environment and on the social channel, and selection on plasticity is measured after each.

The extended synthesis also widens inheritance to include information
transmitted socially \cite{danchin11,laland15}, which makes social learning
the case in which plasticity and inheritance meet. Its evolution has a mature
theory. Boyd and Richerson \cite{boyd85} formalised the biases by which
individuals choose what to copy, including the conformity bias that the
second parameter of our social channel borrows; Rogers \cite{rogers} showed
that social learners can invade a population without raising its mean
fitness, because they consume information that others produce; Borenstein,
Feldman and Aoki \cite{borenstein08} and Ram, Liberman and Feldman
\cite{ram18} showed that which mode of learning selection favours depends on
how the environment fluctuates, and that the stable rate of a transmission
mode can differ from the rate that maximises population fitness; Aoki and
Feldman \cite{aokifeldman14} review the field. These models are analytic and
precise, but their variable is the frequency of a learning strategy, not
selection on plasticity within a genome, so they cannot say whether a social
channel instructs or permits. On mechanism, comparative psychology and
neuroscience agree: social and asocial learning use the same learning
process, and what is distinctively social is the input, perceptual and
attentional processes tuned towards other individuals
\cite{heyes12,heyes,olsson}. This paper adopts that architecture, one learning rule with a heritable social input, which is what allows selection on the input itself to be measured.

Models of learning and evolution supply the setting in which plasticity and
its assimilation can be counted exactly. Hinton and Nowlan \cite{hn} (henceforth H\&N) showed
that lifetime learning lets a population find a target that genetic search
alone would not, and the model has been analysed repeatedly: Fontanari and
Meir \cite{fontanari} and Belew \cite{belew} derived its waiting times,
Harvey \cite{harvey} showed that the plasticity it retains is maintained by
drift rather than by selection, Ancel \cite{ancel00} showed in a
quantitative-genetic version that plasticity can even slow adaptation,
Mayley \cite{mayley} added a cost of learning, and Morgan et al.\ \cite{morgan} and Santos et al.\ \cite{santos}
warned that what the effect shows depends on the kind of plasticity assumed.
Nolfi and colleagues reproduced the effect in neural networks and noted that
it raised average rather than peak fitness \cite{nolfi94,nolfi96}. Social
learning was added in three forms that differ in where the social signal
enters the learner: as the learner's training target
\cite{hutchins,denaro96,curran07}, as a bias on the learner's own guesses
\cite{belew}, or as a value copied directly into its undetermined loci,
bypassing learning
\cite{best,le19,le19gecco,le20evoapps,heinerman,heinerman2,bredeche}; recent
neural-network \cite{chisausky25} and reinforcement-learning \cite{cook24}
models use the first two. In every case the evidence that social learning
favours plasticity is the plasticity remaining at the end of a run, and the
tradition carries an unresolved contradiction, between Best's \cite{best}
report that social learning alone can guide a population to its target and
Le's \cite{le19} that it cannot, a contradiction resolved, with the present model in hand, in Appendix~\ref{app:bestle} and taken up in the Discussion.

The instruments the question needs already exist. Lande and Arnold
\cite{lande83} showed that selection on a trait can be measured as the
regression of relative fitness on the trait within a generation, with no
assumption about inheritance, and the method has since been applied to
thousands of traits in natural populations \cite{kingsolver01}; it has not
been applied to plasticity in the models above. Harvey \cite{harvey}
supplied the complementary test for drift on this very landscape, the
distribution of per-locus frequencies, which separates a trait held at an
intermediate frequency by selection from one that persists because
selection has stopped. With a selection gradient on plasticity, a convergence
test and a drift test applied to one model in which plasticity is a count,
the causal question above becomes answerable, and the next section builds
that model.

\section{Model}
\label{sec:model}

The model has three parts: a population whose members carry some traits
fixed by inheritance and leave others to be learned; a social channel through
which the learning of one individual is informed by others; and an
environment that may change. We define each in turn, then derive what the
model implies about selection on plasticity before any simulation is run,
and finally describe how selection is measured in simulation.

\subsection{Choice of landscape}

We place the model on the landscape of H\&N for three reasons. First,
plasticity is defined exactly on it: each locus is either fixed by
inheritance or left open, so an individual's plasticity is a single count
and selection on it can be measured without ambiguity. Second, the landscape
has no fitness gradient other than the one learning creates, so any selection
on plasticity that we measure is due to learning and the environment, not to
the structure of the fitness function. Third, it is the landscape on which
the computational claims reviewed in Section~\ref{sec:related} were made, so a
result on it speaks to those claims directly.

\subsection{Population and lifetime learning}

A population of $N$ haploid individuals is followed in discrete, non-overlapping
 generations. Each individual $i$ carries $L$ loci, and the allele
$x_{ij}$ at locus $j$ takes one of three values: $0$ or $1$, which are
\emph{determined} by inheritance, or ``?'', which is \emph{undetermined} and
must be set by learning. The environment rewards a single target string
$\theta = (\theta_1,\dots,\theta_L) \in \{0,1\}^L$. We call the number of
undetermined loci of an individual its \emph{plasticity},
\[
n_{?,i} \;=\; \#\{\,j : x_{ij} = {?}\,\},
\]
and write $q$ for a particular value of $n_?$. The population frequency of
undetermined alleles, the mean of $n_{?,i}/L$ over individuals, is written
$p(?)$ throughout.

An individual with a determined allele that differs from the target at any
locus cannot express the target and receives the minimum fitness
$F_{\min} = 1$. An individual whose determined alleles all match the target
learns by trial and error: in each of up to $G$ trials it sets every
undetermined locus at random, as described below, and it stops at the first
trial $g$ on which all of them match the target. Its fitness is
\begin{equation}
F \;=\; 1 + 19\,\frac{G - \min(g, G)}{G},
\label{eq:fit}
\end{equation}
which runs from $F_{\min} = 1$ (no success within $G$ trials) to
$F_{\max} = 20$ (an individual with no undetermined loci, which succeeds
without learning). The earlier an individual learns the target, the fitter it
is, so undetermined loci carry a cost: each one is another locus that must be
guessed correctly.

Parents are drawn with probability proportional to fitness, and offspring
are formed by single-point crossover of two parents. There is no mutation at
the loci, so the only sources of change in the genetic composition of the
population are selection and drift.

\subsection{The social channel}
\label{sec:model_split}

Learning is informed by the previous generation. Let the
\emph{demonstrators} be the individuals of the previous generation that
reached the target ($F > F_{\min}$). A learner observes the phenotype of $k$
demonstrators, sampled with replacement; at each locus a demonstrator
expresses the target value, because it reached the target. Each observed value
is perceived correctly with probability $f$, the \emph{fidelity} of
transmission, and as the opposite value otherwise. The \emph{social cue} at
locus $j$, $c_j \in [0,1]$, is the fraction of the $k$ perceived values that
equal $1$.

The learner combines the cue with its own uninformed guess in a single
learning rule. Its probability of setting locus $j$ to $1$ on a trial is
\begin{equation}
\pi_j \;=\; D_b(\hat c_j),
\qquad
\hat c_j \;=\; a\,c_j + (1-a)\,\tfrac12,
\label{eq:periph}
\end{equation}
where
\begin{equation}
D_b(x) \;=\; x + b\,x(1-x)(2x-1).
\label{eq:conform}
\end{equation}
The two parameters correspond to the two parts of the architecture of
Section~\ref{sec:related}. The \emph{attention gate} $a \in [0,1]$ sets how
much social information enters the learning rule: at $a = 0$ the cue is
ignored and every guess is a fair coin, which is the asocial model of H\&N;
at $a = 1$ the learner guesses in proportion to the cue. The
\emph{conformity transform} $D_b$, with $b \in [-1,1]$, sets how the admitted
information is processed. It fixes the points $0$, $\tfrac12$ and $1$ and
satisfies $D_b(1-x) = 1 - D_b(x)$; $b > 0$ pushes a cue towards the majority
value, $b < 0$ pulls it towards $\tfrac12$, and $b = 0$ uses it as it is. The
social channel therefore changes the input to learning, not the learning
rule: guessing is the same process with or without it.

How demonstrators are sampled turns out to matter. Unless stated otherwise, the $k = 10$ demonstrators are drawn uniformly from all those of
the previous generation. A second rule is used in
Sections~\ref{sec:r7} and~\ref{sec:r6}: $k = 5$ demonstrators drawn from the
fittest tenth of the previous generation, so that the cue is itself a sample
that varies from one generation to the next. A third, a single demonstrator
that is the fitter of the learner's two parents, so that the cue is
inherited along a lineage and carries no variance of its own, is used in Appendix~\ref{app:bestle}.

In most experiments $a$ and $b$ are set by the experimenter and are the same
for every individual, so that they can be intervened on. In
Section~\ref{sec:r6}, $a$ is instead a heritable trait: each offspring
inherits the value of one parent with a Gaussian perturbation, so that
selection on the channel itself can be observed.

\subsection{Environmental change}

Each generation, with probability $\psi$, one locus of the target, chosen at
random, is flipped. The rate $\psi$ is the environmental parameter of the
model; $\psi = 0$ is a static environment. A flip turns every individual that
carries a determined allele at that locus from a match into a mismatch, and
leaves individuals for whom the locus is undetermined unaffected, since they
set it afresh by learning. Because the loci do not mutate, a population in
which no individual any longer reaches the target cannot recover it; we say
such a population has \emph{collapsed}, and write $\Pr(\mathrm{live})$ for
the fraction of runs that still contain target-reaching individuals (more
than one-half of a winner per generation, averaged over the final fifty
generations). A collapsed population persists at the fitness floor; nothing
demographic is implied.

\subsection{Analysis: where the environment and the channel enter selection}
\label{sec:model_sign}

We now derive, for this model, which parameters can change the direction of
selection on plasticity and which can only change its strength. The
derivation uses a mean-field approximation: the cue is replaced by its
expectation, and a single generation is considered.

\paragraph{Probability of learning the target.} Because a demonstrator
expresses the target value at every locus, the expected fraction of
perceived values that match the target is $f$. By the symmetry
$D_b(1-x) = 1 - D_b(x)$, the probability that a learner sets locus $j$ to its
target value is then the same whether $\theta_j = 0$ or $1$:
\begin{equation}
p \;=\; D_b\bigl(a f + (1-a)\,\tfrac12\bigr).
\label{eq:p}
\end{equation}
Guesses at different loci and on different trials are independent, so an
individual whose $q$ undetermined loci are the only ones it must learn
succeeds on a given trial with probability $P = p^{\,q}$, and the number of
trials it uses is geometric, truncated at $G$:
\begin{equation}
E\bigl[\min(g,G)\bigr] \;=\; \sum_{t=0}^{G-1} (1-P)^t \;=\;
\frac{1-(1-P)^G}{P}.
\label{eq:chain}
\end{equation}
By equation~\eqref{eq:fit}, the expected fitness of an individual whose
determined alleles all match the target is
\begin{equation}
F_{\mathrm{learn}}(q) \;=\; 1 + 19\,\Bigl(1 - \frac{1-(1-p^{q})^G}{G\,p^{q}}\Bigr).
\label{eq:flearn}
\end{equation}
This depends on $a$ and $b$ only through $p$, and it does not depend on $\psi$.

\paragraph{Selection on plasticity.} We measure selection on plasticity as the
regression of fitness on plasticity across the individuals of a
generation \cite{lande83},
\begin{equation}
\beta_q \;=\; \frac{\operatorname{Cov}(F, n_?)}{\operatorname{Var}(n_?)}.
\label{eq:betaq}
\end{equation}
Since $\operatorname{Cov}(F, n_?) = \operatorname{Cov}(E[F \mid n_?], n_?)$,
$\beta_q$ is the regression of the conditional mean fitness $E[F \mid q]$ on
$q$. If $E[F \mid q]$ changes in the same direction at every value of $q$
present in the population, $\beta_q$ has the sign of
$\Delta(q) = E[F \mid q+1] - E[F \mid q]$, the change in expected fitness
when one more locus is left undetermined. Positive $\beta_q$ means that
selection favours plasticity, negative that it removes it, and zero that it
is neutral.

\paragraph{The static environment.} When $\psi = 0$, and once determined
alleles that do not match the target have been removed, every individual's
expected fitness is $F_{\mathrm{learn}}(n_?)$.

\begin{proposition}\label{prop:static}
If $\psi = 0$ and all determined alleles match the target, then
$\Delta(q) \le 0$ for every $q$ and every value of $a$, $b$ and $f$, with
equality only if $p = 1$.
\end{proposition}

\noindent\emph{Proof.} Leaving one more locus undetermined multiplies the
per-trial success probability by $p \le 1$, from $p^{q}$ to $p^{q+1}$. The
expected number of trials in equation~\eqref{eq:chain} is a sum of terms
$(1-P)^t$, each non-increasing in $P$, so it does not decrease; by
equation~\eqref{eq:fit}, expected fitness does not increase. If $p < 1$ the
decrease is strict. \hfill$\square$

\smallskip
The social channel enters this result only by raising $p$ towards $1$. It
can therefore make the cost of plasticity arbitrarily small, but it cannot
make plasticity advantageous. Before mismatching determined alleles have been
removed, an undetermined locus can be favoured, because a determined one may
be wrong; this is the classical Baldwin effect on this landscape, and it is
transient in a static environment. The next case shows that a changing
environment makes the same mechanism permanent.

\paragraph{The changing environment.} When $\psi > 0$, a determined allele
can become a mismatch. With one locus flipped per change event, an
individual with $q$ undetermined loci has a determined allele at the flipped
locus with probability $(L-q)/L$, so its target is spoiled in a given
generation with probability $\psi(L-q)/L$, after which its fitness is
$F_{\min}$. Its expected fitness is therefore
\begin{equation}
E[F \mid q] \;=\; \Bigl(1 - \psi\,\frac{L-q}{L}\Bigr)\,F_{\mathrm{learn}}(q)
\;+\; \psi\,\frac{L-q}{L}\,F_{\min}.
\label{eq:ef}
\end{equation}
Write $C(q) = F_{\mathrm{learn}}(q) - F_{\mathrm{learn}}(q+1) \ge 0$ for the
cost of one more undetermined locus. Subtracting equation~\eqref{eq:ef} at
$q$ from its value at $q+1$ and collecting terms gives
\begin{equation}
\Delta(q) \;=\; \underbrace{\frac{\psi}{L}\,
\bigl(F_{\mathrm{learn}}(q+1) - F_{\min}\bigr)}_{\text{benefit}}
\;-\; \underbrace{\Bigl(1 - \psi\,\frac{L-q}{L}\Bigr)\,C(q)}_{\text{cost}}.
\label{eq:delta}
\end{equation}

\begin{proposition}\label{prop:moving}
The benefit of plasticity in equation~\eqref{eq:delta} is proportional to
the rate of environmental change $\psi$ and is zero when $\psi = 0$. The
parameters of the social channel, $a$ and $b$, enter $\Delta(q)$ only through
$p$, and therefore only through $F_{\mathrm{learn}}$ and the cost $C(q)$.
Selection favours plasticity at $q$ if and only if
\begin{equation}
\psi \;\ge\; \psi^{\ast}(q) \;=\;
\frac{L\,C(q)}{F_{\mathrm{learn}}(q+1) - F_{\min} + (L-q)\,C(q)}.
\label{eq:psistar}
\end{equation}
\end{proposition}

\noindent\emph{Proof.} The first two statements follow from
equations~\eqref{eq:p}--\eqref{eq:ef}, in which $\psi$ appears only in the
probability of being spoiled and $a$, $b$ only in $p$. Setting
$\Delta(q) \ge 0$ in equation~\eqref{eq:delta} and solving for $\psi$ gives
equation~\eqref{eq:psistar}. \hfill$\square$

\smallskip
Numerically, $\psi^{\ast}(q)$ decreases monotonically as $p$ increases, at
every $q$ from $1$ to $L-1$ (for $L = 20$, $G = 1000$), and tends to zero as
$p \to 1$, because the cost $C(q)$ vanishes. A more open channel therefore
lowers the rate of change at which plasticity becomes favoured.

Equation~\eqref{eq:ef} also idealises the cue: in simulation the cue is
built from the previous generation's winners, so in the generation after a
flip it points at the old value of the flipped locus, a cost of relying on
the channel that grows with $\psi$ and that the derivation omits. The
omission biases the predicted threshold downwards, and consistently, every
observed crossing in Section~\ref{sec:r45} lies at or above its
prediction.

\paragraph{Causal roles.} Equation~\eqref{eq:delta} assigns the two
parameters different jobs, and we read their causal roles off that
decomposition, in the sense of causal specificity \cite{woodward10,calcott}
applied term by term. The environment is the \emph{instructive} cause of
selection for plasticity: the benefit term exists only through $\psi$ and
vanishes identically at $\psi = 0$ (Proposition~\ref{prop:static}), whatever
the channel does. The channel is \emph{permissive}: $a$ and $b$ enter only
the cost term, so opening the channel rescales the penalty and relocates the
threshold $\psi^{\ast}$ (Proposition~\ref{prop:moving}), but cannot create a
benefit where the environment supplies none. The asymmetry is between the
terms, not between which knob can flip an observed sign: at a fixed
$\psi > 0$, opening the channel can tip $\Delta$ from negative to positive
by cheapening plasticity, just as removing a constraint can expose a
selection pressure it was masking. What no setting of the channel can do is
make plasticity favoured in a world that never changes. The channel also
matters in a way equation~\eqref{eq:delta} does not capture: whether a
population survives repeated changes long enough for selection to act at
all. That is a population-level property, and it is measured in
Section~\ref{sec:results}.

The analysis is a one-generation, mean-field approximation evaluated at a
single value of $q$. It predicts the sign of selection where $\Delta(q)$ is
clearly positive or negative across the population, and may fail near the
threshold, where the population spans values of $q$ on both sides. Whether it
predicts the simulation is tested in Section~\ref{sec:r45}.

\subsection{Measurement in simulation}
\label{sec:measure}

Selection on plasticity is computed in every generation as $\beta_q$ in
equation~\eqref{eq:betaq}, in units of fitness per undetermined locus. Two
summaries are reported. The first averages $\beta_q$ over the final
generations of a run. Under a changing environment plasticity can approach
fixation, so that $\operatorname{Var}(n_?)$ becomes small and $\beta_q$
noisy; we then also report $\beta_q \cdot \mathrm{SD}(n_?)$, the expected
difference in fitness between individuals one standard deviation apart in
plasticity, and use a second, more conservative summary: the average of
$\beta_q$ over every generation after a burn-in of $50$ in which
$\operatorname{Var}(n_?) \ge 0.05$. Each surviving run then contributes one
estimate, and we report how many runs give a positive estimate, with an exact two-sided binomial test against one half.
All values of $\beta_q$ are conditional on the population retaining the
target; runs that fix completely at ``?'' contribute no end-of-run estimate,
and the run counts reported alongside each cell carry the between-run
variation. With $84$ cells in the sweep we treat individual cell tests as
descriptive; the only cell-level result claimed as confirmatory is the
unanimity of the $\psi=0$ column, whose per-cell $p = 4.9\times10^{-4}$
survives a Bonferroni correction across the grid. When $a$ is heritable
(Section~\ref{sec:r6}), the same regression with $a$ in place of $n_?$
measures selection on the channel itself. All simulations reported here
were run with NumPy 2.4.3 under fixed random seeds; random streams differ between
NumPy versions, so exact reproduction requires the archived environment.

Two further tests separate selection from drift. In a \emph{convergence
test}, runs are started from different initial levels of plasticity; if
selection holds plasticity at an equilibrium, the runs converge on it, and
if it does not, they retain their initial differences. In a \emph{locus
frequency test} \cite{harvey}, we record the frequency of ``?'' at each
locus at the end of long runs: a level of plasticity maintained by selection
appears as loci held at intermediate frequencies, whereas plasticity that
persists because selection no longer removes it appears as loci fixed at
either end, with the population mean equal to the fraction of loci fixed at
``?''.

\section{Results}
\label{sec:results}

Unless stated otherwise: $N=1000$, $L=20$, $G=1000$, $400$ generations,
$k=10$ demonstrators drawn uniformly from those of the previous generation,
fidelity $f=0.9$, $a$ and $b$ fixed and equal for all individuals, one locus
flipped per change event, and $12$ independent runs per condition; in the
sweep of Section~\ref{sec:r45} one demonstrator sample is drawn per
generation and shared by all learners. Reported values are means over the
final generations of each run (the convergence runs of Table~\ref{tab:r2}
last $600$ generations). Sections~\ref{sec:r7} and~\ref{sec:r6} use a second
implementation of the same model: learning budget $G=300$, $k=5$
demonstrators sampled per learner from the fittest tenth of the previous
generation (winners or not; a non-winner displays its current guesses at
undetermined loci), perfect transmission ($f=1$), $b=0$, and initial
$p(?)=0.5$. Absolute fitness values are therefore not comparable between the
two implementations, and no cross-implementation comparison is made. Each
subsection asks one question of the analysis of Section~\ref{sec:model_sign}
and reports the measurement that answers it.

\subsection{Does any setting of the channel make selection favour plasticity
in a static environment?}
\label{sec:r1}

Proposition~\ref{prop:static} says no: at $\psi = 0$, once mismatched
determined alleles are gone, one more undetermined locus can only slow
learning, whatever the channel admits or how it processes it. What the
analysis does not say is how much of the penalty the channel removes, and
what the plasticity that escapes removal then does; those are measurements.
Table~\ref{tab:r1} gives $\beta_q$ and the final frequency of undetermined
alleles for three configurations, each started from a low and a high initial
level of plasticity.

\begin{table}[htbp]\centering\small
\caption{Selection on plasticity in a static environment ($\psi=0$).
$\beta_q$ is fitness per plastic allele, equation~\eqref{eq:betaq}.}
\label{tab:r1}
\begin{tabular}{@{}lrrrr@{}}
\toprule
& \multicolumn{2}{c}{initial $p(?)=0.2$} & \multicolumn{2}{c}{initial $p(?)=0.8$}\\
\cmidrule(lr){2-3}\cmidrule(lr){4-5}
configuration & $\beta_q$ & final $p(?)$ & $\beta_q$ & final $p(?)$\\
\midrule
$a=0$ (asocial, original model) & $-0.2106$ & $0.1661$ & $-0.1132$ & $0.1362$\\
$a=0.8$, $b=0$ (attention only) & $-0.0167$ & $0.3319$ & $-0.0625$ & $0.6350$\\
$a=0.8$, $b=0.5$ (attention and conformity) & $-0.0085$ & $0.3271$ & $-0.0386$ & $0.8001$\\
\bottomrule
\end{tabular}
\end{table}

$\beta_q$ is negative in every cell. Opening the attention gate weakens the
penalty on plasticity by a factor of $12.6$, and adding conformity weakens it
by $24.8$, but neither changes its sign. On the standard diagnostic the same
runs would look like a strong effect: the final frequency of undetermined
alleles rises from $0.17$ to $0.33$ and, from a high start, to $0.80$. Read
through $\beta_q$, nothing has been selected for. The social channel has made
plasticity cheap, not advantageous.

If the weakened penalty of Table~\ref{tab:r1} were instead a selective
equilibrium at a higher level of plasticity, runs started at different
levels would converge on it.  Table~\ref{tab:r2}
starts three runs at $p(?)=0.2$, $0.5$ and $0.8$, an initial spread of $0.60$,
and reports what remains.

\begin{table}[htbp]\centering\small
\caption{Convergence test. Final spread is the range across the three starting
points; the last column expresses it as a fraction of the initial spread of
$0.60$.}
\label{tab:r2}
\begin{tabular}{@{}lrrrrr@{}}
\toprule
configuration & from $0.2$ & from $0.5$ & from $0.8$ & final spread & initial spread retained\\
\midrule
$a=0$ (asocial) & $0.1591$ & $0.1008$ & $0.1199$ & $0.0583$ & $9.7\%$\\
$a=0.8$ & $0.3332$ & $0.4820$ & $0.5663$ & $0.2331$ & $38.9\%$\\
$a=1.0$ & $0.3047$ & $0.5769$ & $0.8321$ & $0.5274$ & $87.9\%$\\
\bottomrule
\end{tabular}
\end{table}

The asocial model erases ninety per cent of its initial condition: selection
is working and there is an equilibrium to find. At $a = 0.8$ the spread is
still closing, from $0.48$ at generation $100$ to $0.23$ at $600$, at a pace
consistent with selection weakened twenty-five-fold rather than removed; a
fixed-horizon spread cannot by itself distinguish neutralised from slowed
selection. At $a = 1.0$ nothing closes, and the final level of plasticity is
a memory of the initial condition. The locus-frequency test of
Section~\ref{sec:r7} makes the distinction cleanly.

The answer to the question of this subsection is therefore no: selection
against plasticity keeps its sign at every opening of the channel; the
channel changes how fast, and ultimately whether, that selection still
moves the population.

The sweep of Section~\ref{sec:r45} closes the remaining gap: in its
$\psi = 0$ column, $\beta_q < 0$ in every one of its $132$ distinct runs
($b$ has no effect at $a = 0$, so two cells coincide), so the answer does
not change at any opening of the channel.

\subsection{What reverses the sign of selection: the channel or the
environment?}
\label{sec:r45}

\begin{figure}[htbp]\centering
\includegraphics[width=\textwidth]{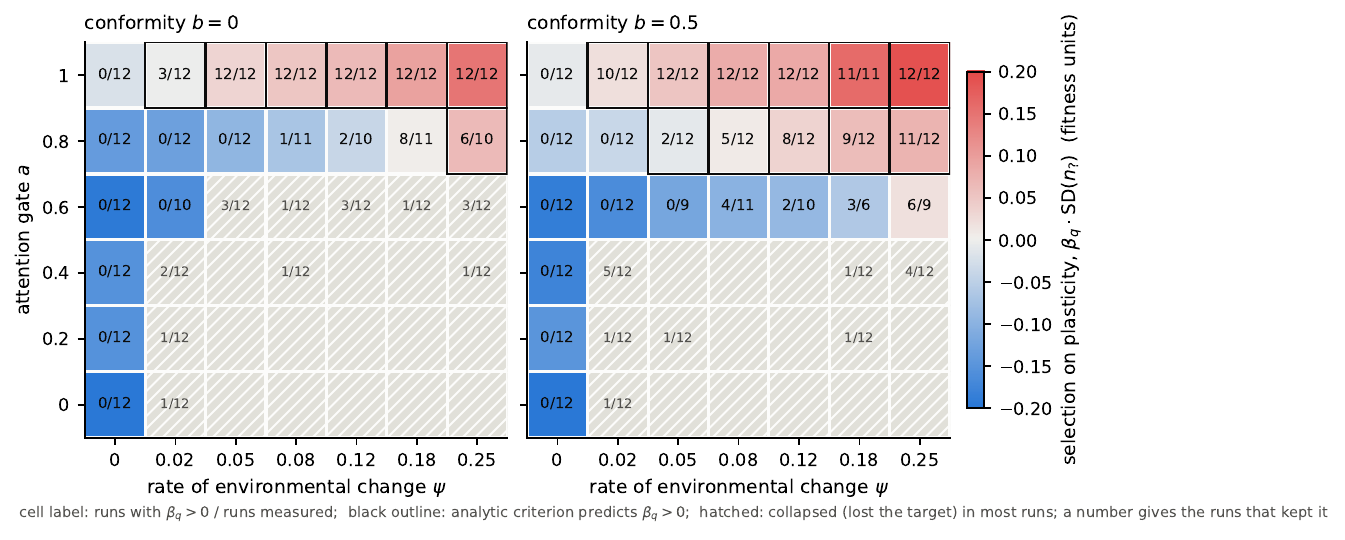}
\caption{Sign of selection on plasticity across channel strength $a$ and
rate of environmental change $\psi$, for conformity $b = 0$ (left) and
$b = 0.5$ (right); $12$ independent runs per cell, initial $p(?) = 0.8$. Cell colour is
the mean of $\beta_q \cdot \mathrm{SD}(n_?)$ over runs, measured along the
trajectory; the label is the number of runs with $\beta_q > 0$ over the
number measured. Hatched cells are
conditions in which the population collapsed (lost the target) in more than
half of the runs. A black outline marks cells in which
equation~\eqref{eq:delta}, evaluated at the cell's own mean plasticity,
predicts $\beta_q > 0$.}
\label{fig:sign}
\end{figure}

Figure~\ref{fig:sign} sweeps the attention gate from $a = 0$ to $1$ and the
rate of change from $\psi = 0$ to $0.25$, for $b = 0$ and $b = 0.5$, twelve
independent runs per cell, with the trajectory estimator of
Section~\ref{sec:measure}. Proposition~\ref{prop:moving} makes three predictions about this plane:
the column $\psi = 0$ is negative at every $a$ and $b$; for an open channel
the sign crosses at $\psi^{\ast}(\bar q)$ of equation~\eqref{eq:psistar};
and the crossing moves to smaller $\psi$ as the channel opens, because the
channel sets only the cost term. The figure shows all three, and one thing
the proposition is silent about.

First, the column $\psi = 0$ is negative at every level of the channel. In
all $132$ distinct runs of that column ($b$ is inert at $a=0$, so two cells
coincide), $\beta_q < 0$, with an exact binomial $p \le 0.001$ in every
cell; the cell
means run from $-0.237$ at $a = 0$ to $-0.007$ at $a = 1$, $b = 0.5$. This is
Proposition~\ref{prop:static} again, now at every level of the channel
rather than at the three of Table~\ref{tab:r1}.

Second, and this is the part equation~\eqref{eq:delta} does not contain,
below $a = 0.6$ populations with $\psi \ge 0.05$ collapse in at least eight
of twelve runs, and in most of those conditions in all twelve; at $a = 0.6$
survival is partial and the majority of runs stays non-positive below
$\psi = 0.25$. The channel decides whether there is a population left to
measure.

Third, for $a \ge 0.8$ the sign reverses with $\psi$, and the crossing moves
to lower $\psi$ as the channel opens. At $a = 1$ with $b = 0$, $\beta_q > 0$
in twelve of twelve runs from $\psi = 0.05$ ($p < 0.001$); with $b = 0.5$
the majority is already positive at $\psi = 0.02$ (ten of twelve,
$p = 0.039$) and unanimous from $\psi = 0.05$; at $a = 0.8$ the majority of runs turns
positive between $\psi = 0.12$ and $0.18$ ($b = 0$) and between $0.08$ and
$0.12$ ($b = 0.5$); at $a = 0.6$ it does not turn positive below
$\psi = 0.25$ for $b = 0$ and does so only at $\psi = 0.25$ for $b = 0.5$.

\begin{table}[htbp]\centering\small
\caption{Predicted and observed thresholds. $p$ is the per-locus probability
of guessing correctly at a cue of accuracy $f = 0.9$; $\bar q$ is the mean
number of undetermined loci in the $\psi = 0.05$ cell, at which
$\psi^{\ast}$ from equation~\eqref{eq:psistar} is evaluated; the observed bracket
is the pair of adjacent grid values between which the majority sign of
$\beta_q$ changes.}
\label{tab:psistar}
\begin{tabular}{@{}rrrrrl@{}}
\toprule
$a$ & $b$ & $p$ & $\bar q$ & predicted $\psi^{\ast}$ & observed crossing\\
\midrule
$0.6$ & $0$   & $0.74$ & $12$ & $0.247$ & none below $0.25$\\
$0.8$ & $0$   & $0.82$ & $16$ & $0.106$ & $0.12$--$0.18$\\
$1.0$ & $0$   & $0.90$ & $18$ & $0.015$ & $0.02$--$0.05$\\
$0.6$ & $0.5$ & $0.79$ & $15$ & $0.200$ & $0.18$--$0.25$\\
$0.8$ & $0.5$ & $0.87$ & $18$ & $0.040$ & $0.08$--$0.12$\\
$1.0$ & $0.5$ & $0.94$ & $19$ & $0.005$ & $0$--$0.02$\\
\bottomrule
\end{tabular}
\end{table}

Table~\ref{tab:psistar} compares these crossings with $\psi^{\ast}(\bar q)$ from
equation~\eqref{eq:psistar}. The predicted threshold lies within or
adjacent to the observed bracket in five of the six configurations and one
grid step early in the sixth ($a = 0.8$, $b = 0.5$). Cell by cell, the sign
of $\Delta(\bar q)$ in equation~\eqref{eq:delta} at the cell's own $\bar q$ matches the
majority sign of $\beta_q$ in $38$ of the $43$ cells with at least six
measured runs, and all five mismatches lie on the boundary, where the
fraction of positive runs is between $2/12$ and $8/11$. The end-of-run
estimate of Table~\ref{tab:r3} put the crossing for $a = 0.8$, $b = 0.5$ at
$\psi \approx 0.06$; the trajectory estimate is the more conservative of the
two and lies in the same region.

The three features are what Propositions~\ref{prop:static}
and~\ref{prop:moving} say. The environment is
the only term that rewards plasticity, so without environmental change the
channel can only weaken the penalty; the channel is the only term that prices
plasticity, so it sets both the threshold at which change makes plasticity
worth having and whether the population survives to reach it, without ever
supplying the direction.

A finer slice through this plane, at the single setting $a=0.8$, $b=0.5$
and with the end-of-run estimator, shows the crossing itself.
Table~\ref{tab:r3} sweeps the rate of environmental change at the
configuration that was most permissive in the static case.

\begin{table}[htbp]\centering\small
\caption{Moving target, $a=0.8$, $b=0.5$, initial $p(?)=0.8$.
$\Pr(\text{live})$ is the fraction of runs whose population survives to the
end. The last column is $\beta_q$ multiplied by the standard deviation of
$n_?$, that is, the expected fitness difference for a one-standard-deviation
difference in plasticity, on a scale from $1$ to $20$.}
\label{tab:r3}
\begin{tabular}{@{}rrrrrr@{}}
\toprule
$\psi$ & $\Pr(\text{live})$ & $p(?)$ & $\beta_q$ & $\operatorname{Var}(n_?)$ & $\beta_q\!\cdot\!\mathrm{SD}$\\
\midrule
$0.00$ & $1.00$ & $0.8001$ & $-0.0386$ & $1.892$ & $-0.0531$\\
$0.02$ & $1.00$ & $0.8349$ & $-0.0188$ & $1.540$ & $-0.0233$\\
$0.05$ & $1.00$ & $0.9132$ & $-0.0163$ & $0.825$ & $-0.0148$\\
$\mathbf{0.08}$ & $1.00$ & $0.9259$ & $\mathbf{+0.0204}$ & $0.688$ & $+0.0169$\\
$0.12$ & $1.00$ & $0.9543$ & $+0.1877$ & $0.358$ & $+0.1123$\\
$0.18$ & $1.00$ & $0.9850$ & $+1.6526$ & $0.121$ & $+0.5739$\\
$0.25$ & $1.00$ & $0.9900$ & $+1.3071$ & $0.078$ & $+0.3640$\\
\bottomrule
\end{tabular}
\end{table}

Linear interpolation puts the sign change at $\psi^\ast=0.0633$. The obvious
objection is that $\operatorname{Var}(n_?)$ falls by a factor of twenty-four
across the sweep, so a large $\beta_q$ could be a small denominator. The final
column answers it: $\beta_q\cdot\mathrm{SD}(n_?)$ changes sign at the same
place and reaches $+0.574$, more than half a fitness unit, so the reversal is
not an artefact of scaling.

The same slice with the channel closed shows what the channel's survival
role means in practice.
Table~\ref{tab:r4} repeats the sweep with the attention gate closed.

\begin{table}[htbp]\centering\small
\caption{Control: the same sweep with $a=0$. Dashes indicate that no
run retained the target, so no coefficient is defined. Bold in
Table~\ref{tab:r3} marks the first row with $\beta_q>0$.}
\label{tab:r4}
\begin{tabular}{@{}rrrrr@{}}
\toprule
$\psi$ & $\Pr(\text{live})$ & $p(?)$ & $\beta_q$ & $\beta_q\!\cdot\!\mathrm{SD}$\\
\midrule
$0.000$ & $1.00$ & $0.1362$ & $-0.1132$ & $-0.0818$\\
$0.005$ & $0.25$ & $0.1781$ & $-0.2022$ & $-0.1619$\\
$0.010$ & $0.00$ & -- & -- & --\\
$0.020$ & $0.08$ & $0.2004$ & $-0.2886$ & $-0.0324$\\
$0.030$ & $0.00$ & -- & -- & --\\
$0.050$ & $0.00$ & -- & -- & --\\
\bottomrule
\end{tabular}
\end{table}

There is no regime in which plasticity is positively selected, because the
population does not survive long enough to have one. Survival has fallen to a
quarter of runs by $\psi=0.005$, and wherever a coefficient can be computed it
is not merely negative but more negative than in the static case. The regime
identified in Table~\ref{tab:r3} is therefore not a property of environmental
change; it is a property of environmental change \emph{together with} a social
channel.

Table~\ref{tab:r5} shows that both components are required, and that the
initial level of plasticity is a third requirement.

\begin{table}[htbp]\centering\small
\caption{Both channels needed, at $\psi=0.05$. Winner is the number of
individuals out of $1000$ expressing the target in the final generation.}
\label{tab:r5}
\begin{tabular}{@{}rrrrrr@{}}
\toprule
$a$ & $b$ & initial $p(?)$ & $\Pr(\text{live})$ & $p(?)\mid$ live & Winner\\
\midrule
$0.0$ & $0.0$ & $0.2$ & $0.00$ & -- & $0$\\
$0.0$ & $0.0$ & $0.8$ & $0.00$ & -- & $0$\\
$0.8$ & $0.0$ & $0.2$ & $0.00$ & -- & $0$\\
$0.8$ & $0.0$ & $0.8$ & $1.00$ & $0.8022$ & $990$\\
$0.8$ & $0.5$ & $0.2$ & $0.08$ & $0.5164$ & $51$\\
$0.8$ & $0.5$ & $0.8$ & $1.00$ & $0.9132$ & $993$\\
\bottomrule
\end{tabular}
\end{table}

With the channel closed the population collapses whatever its plasticity. With the
channel open but little plasticity to begin with, it also collapses. Survival
requires the peripheral gate open \emph{and} a stock of plasticity for the
incoming signal to act on, and conformity, the central parameter, rescues a
small fraction of the otherwise fatal low-plasticity condition rather than
substituting for either.

\subsection{Is the plasticity the channel preserves held by selection, or by
drift?}
\label{sec:r7}

\begin{figure}[htbp]\centering
\includegraphics[width=\textwidth]{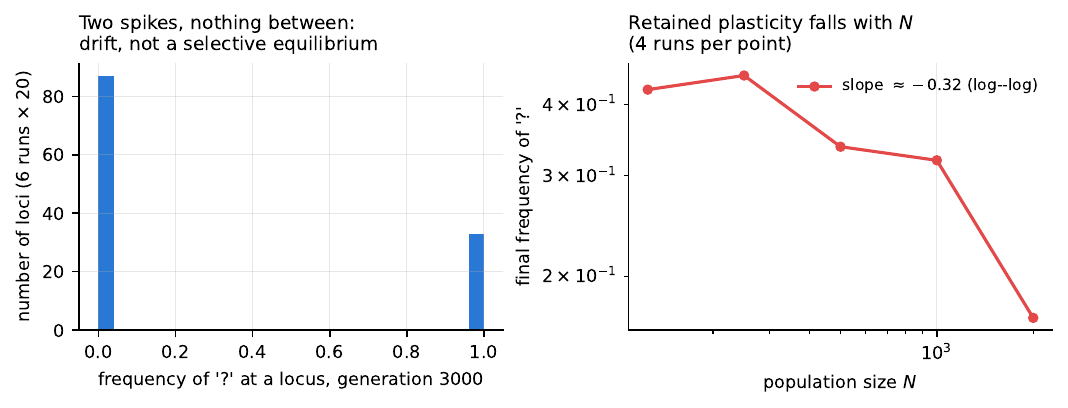}
\caption{Left: the distribution, across loci and six long runs, of the
frequency at which each locus sits after 3000 generations, with $a$ evolving,
$k=5$, fittest tenth. The distribution is two spikes at the ends, with
essentially nothing between. Right: the final population frequency of ``?''
falls with population size $N$, consistent with a drift process rather than a
fixed equilibrium.}
\label{fig:drift}
\end{figure}

Proposition~\ref{prop:static} leaves only one candidate for what maintains
the plasticity that an open channel preserves in a static environment:
since no interior level of plasticity is an optimum, a population can hold
one only by drift. Drift makes three predictions that selection does not.
Individual loci should fix at one end or the other, rather than sit at a
shared intermediate frequency; the population mean should equal the
fraction of loci that happen to have fixed at ``?''; and the retained level
should shrink as the population grows. The locus-frequency test of
Section~\ref{sec:measure} measures the first two directly: a selective equilibrium at an intermediate allele
frequency should show up as a frequency at which loci actually sit, not only
as a stable population mean, which can equally be produced by a population
split between loci fixed at either end. Figure~\ref{fig:drift} reports the result, and Table~\ref{tab:drift} collects the three fractions.

\begin{table}[htbp]\centering\small
\caption{Locus-frequency distribution after $T=3000$ generations, six runs
$\times\,20$ loci, $a$ evolving, $k=5$, fittest tenth. Bold marks the stuck
class, whose fraction equals the population frequency of ``?''.}
\label{tab:drift}
\begin{tabular}{@{}lr@{}}
\toprule
loci fixed to a determined value (frequency $<0.01$) & $0.725$\\
loci \textbf{stuck} at ``?'' (frequency $>0.99$) & $\mathbf{0.275}$\\
loci at an intermediate frequency ($0.15$--$0.85$) & $0.000$\\
\bottomrule
\end{tabular}
\end{table}

The distribution is bimodal: no locus sits at an intermediate frequency
after three thousand generations, which is not what a selective equilibrium
looks like and is exactly what drift looks like. The stuck-locus fraction,
$0.275$, equals the population-level ``?'' frequency of the same runs; the
same configuration stood at $0.327$ at generation 2500 (six runs) and is
still drifting towards fixation at 3000. This is Harvey's
identity, that the population mean under drift equals
the fraction of loci that happen to have fixed in the undetermined state, not
a frequency that any single locus is held at by selection. Best
\cite{best}, looking at the same persistence a quarter-century ago, judged
that ``the forces at hand are genetic drift \ldots{} versus this selection
pressure towards assimilation''; the measurements above give that sentence
its numbers, and this paper its title. The final
``?''-frequency also falls as $N$ grows across a sweep from $N=125$ to
$N=2000$ (noisy at the smaller sizes, clearer for $N\geq500$), the signature of
a quantity shrinking under drift as the population gets larger, which a
selected equilibrium frequency would not do.

All three predictions of drift are met, and none of selection. The social channel does not reward plasticity, it
relaxes the penalty on it, and genetic drift does the rest, settling at a
higher stuck fraction than the asocial model's $0.094$--$0.100$ because the
penalty it is drifting against is weaker. We have not found a configuration,
across Sections~\ref{sec:r1}--\ref{sec:r7}, in which an open social channel on
a static landscape produces a selective reward for plasticity; every case we
can measure is consistent with permission rather than instruction, now by
three independent instruments -- the regression $\beta_q$, the convergence
test, and Harvey's own locus-frequency diagnostic.

\subsection{If the channel is permissive, why does selection open it?}
\label{sec:r6}

\begin{figure}[htbp]\centering
\includegraphics[width=\textwidth]{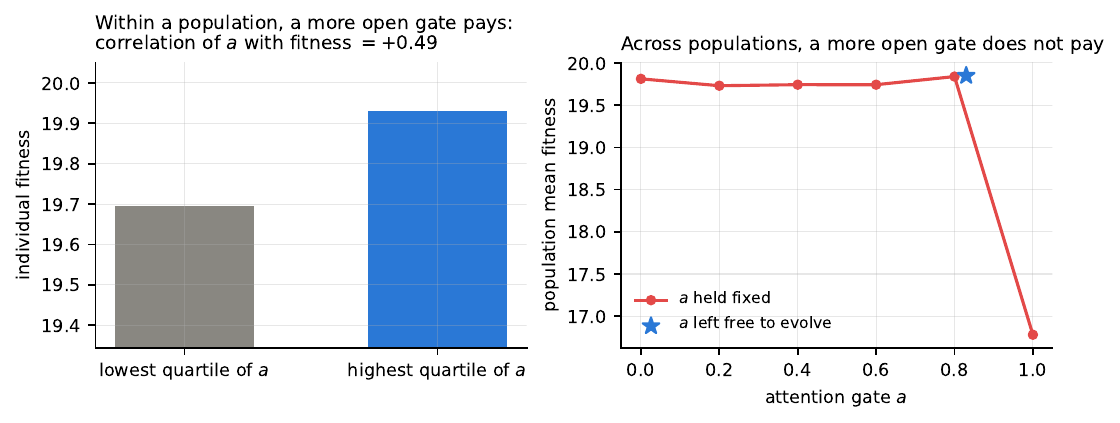}
\caption{Left: within a population with $a$ evolving, individuals with a
more open gate earn more fitness (the vertical axis does not start at
zero). Right: across populations with $a$ held fixed, mean fitness is flat
until the channel is fully open and collapses at $a=1$; the star is the
population left free to evolve $a$.}
\label{fig:rogers}
\end{figure}

Everything so far holds the channel fixed. Equation~\eqref{eq:delta},
however, is a statement about a single individual: whatever $\psi$ is,
raising $a$ lowers that individual's cost of plasticity, so wherever
plasticity exists, an individual with a more open gate should outperform
one with a less open gate, whether or not the population as a whole gains
anything. The analysis therefore predicts a gap between the two levels of
selection, which is Rogers' \cite{rogers} structure. Letting $a$ evolve as
a heritable trait (Section~\ref{sec:measure}) makes both levels measurable
at once (figure~\ref{fig:rogers}).

\begin{table}[htbp]\centering\small
\caption{Individual versus population: $400$ generations, $k=5$, fittest
decile, $a$ evolving with Gaussian mutation, last $100$ generations averaged.}
\label{tab:rogers}
\begin{tabular}{@{}lr@{}}
\toprule
correlation between individual $a$ and individual fitness & $+0.491$\\
mean fitness, top quartile of $a$ & $19.930$\\
mean fitness, bottom quartile of $a$ & $19.694$\\
mean evolved $a$ & $0.830$\\
mean fitness of the population that evolves $a$ freely & $19.848$\\
\bottomrule
\end{tabular}
\end{table}

\begin{table}[htbp]\centering\small
\caption{Population mean fitness with $a$ held fixed, same conditions as
Table~\ref{tab:rogers}. Bold marks the collapse at $a=1$.}
\label{tab:rogers_fixed}
\begin{tabular}{@{}lrrrrrr@{}}
\toprule
$a$ (fixed) & $0.0$ & $0.2$ & $0.4$ & $0.6$ & $0.8$ & $1.0$\\
mean fitness & $19.811$ & $19.730$ & $19.742$ & $19.742$ & $19.838$ & $\mathbf{16.781}$\\
\bottomrule
\end{tabular}
\end{table}

Within the population, the correlation between an individual's own $a$ and its
own fitness is positive and not small: the most socially attentive quarter of
the population outperforms the least attentive quarter by about a fifth of a
fitness point. Selection accordingly pushes $a$ up to $0.83$, short of the
collapse at $a=1$. But holding $a$ fixed across whole populations shows the
opposite at the extreme: a population held at $a=1$ earns three fitness
points less than one held at $a=0$, because a population in which nobody
guesses independently has no new information for the social channel to
carry. This is Rogers' \cite{rogers} paradox -- a trait that is individually
advantageous to adopt can leave the population no better off, or worse off,
than not having it -- and nothing in equation~\eqref{eq:periph} encodes it; it
falls out of letting $a$ evolve on a landscape where information has to be
produced by someone before it can be copied by anyone. The population that
evolves $a$ freely reaches $19.85$, as high as the best fixed setting in
Table~\ref{tab:rogers_fixed}, which is the standard escape from the paradox \cite{enquist07,aokifeldman14}:
such a population stays polymorphic, with low-$a$ individuals continuing to
generate the information that high-$a$ individuals free-ride on, rather than
collapsing onto the individually-favoured extreme.

\section{Discussion}
\label{sec:discussion}

The social channel in this model is permissive, and the analysis says in
what sense. The change in expected fitness when one more locus is left
undetermined, equation~\eqref{eq:delta}, has one term that rewards
plasticity, and it is proportional to the rate of environmental change; the
channel parameters appear only in the term that charges for it. In a static
environment the reward is zero, and no setting of the channel makes
selection favour plasticity: the channel weakens the penalty by more than an
order of magnitude and at full strength neutralises it, but it never
reverses it. When the environment changes, selection on plasticity reverses
sign at a threshold that the channel lowers and cannot cross alone, and
without the channel the population collapses before the threshold is
reached. In the vocabulary of developmental biology, environmental change is
the instructive cause of selection for plasticity and social learning is a
permissive one: necessary, dose-dependent, and silent about direction.

This bears directly on the debate with which we began. The disagreement
between the extended and the standard synthesis over plasticity-led
evolution is in large part a disagreement about causal role, whether
plasticity directs adaptive evolution or only permits it
\cite{laland15,levis16,uller19}. The model does not settle that question for
nature. It shows that the question can be settled by measurement in a case
where it is posed exactly, and it returns an answer that gives each side its
due. The direction of selection on plasticity is set by the environment, as
the standard account holds. The channel through which socially acquired
information arrives is a condition without which that direction is never
expressed, and whose strength sets how much environmental change is needed
before it is, which is what the extended account claims for inclusive
inheritance. Permissive is not dispensable. That environmental change is
what makes plasticity worth paying for is classical quantitative-genetic
ground \cite{ancel00,chevin10,scheiner17}; what the decomposition adds is
where a social channel can, and cannot, enter that account.

The methodological consequence is the sharpest. A high terminal level of
plasticity is compatible with strong positive selection and with no
selection at all, and in this model it is the latter: what an open channel
preserves in a static world is drift under a relaxed penalty, with no locus
held at an intermediate frequency. Any claim of the form ``plasticity was
maintained, therefore plasticity was favoured'' needs a direct measurement
of selection or a convergence test, and both are cheap; the selection
gradient and the drift test used here are the instruments that Levis and
Pfennig \cite{levis16} ask for in natural populations. The same caution
applies retrospectively to the computational literature, including Belew's
\cite{belew} reading of a raised plasticity as selection for learning and
the conclusions of Le and colleagues \cite{le19}. It also dissolves an old
disagreement in that literature: whether social learning alone can guide a
population to its target \cite{best,le19} turns, in the present model, not
on the learning rule but on whether the demonstrator sample carries variance
of its own, a single pool-sampling parameter that neither study varied
(Appendix~\ref{app:bestle}).

Two further readings follow. First, the split of the channel into a
peripheral gate and a central transform gives Heyes's account
\cite{heyes12,heyes} a form a model can test rather than assume, and the
test comes out asymmetric in the way her account expects: the peripheral
parameter is necessary, since closing it is lethal under environmental
change at every level of plasticity we tried, while the central parameter
only deepens the discount and rescues a fraction of otherwise doomed
populations. Second, what survives of the Baldwin effect here is a claim
about persistence, not speed. The canonical reading of H\&N is that learning
finds the target faster. Under environmental change the asocial population
is not slower, it has collapsed; the channel does not accelerate the search so
much as keep a population alive long enough for selection on plasticity to
become positive at all. That is closer to Baldwin's own pairing of organic
selection with social heredity \cite{baldwin} than to the search-efficiency
reading: the means of staying in the game are socially supplied, and the
reasons to change are not.

What the exercise offers the wider debate is less a verdict than a
protocol. Any claim that an inclusive-inheritance channel, social learning,
parental effects, epigenetic marks, leads or merely accompanies adaptive
evolution can be put in the form used here: write expected fitness as a
function of the trait, ask which term of the resulting decomposition the
channel enters, and measure the selection gradient under interventions on
the channel and on the environment separately. Three extensions seem most
informative. First, landscapes with structure, where learning has a gradient
of its own, test whether the permissive verdict is a property of social
information or of the needle. Second, the predicted scaling of the threshold
with $L/G$, and the effect of a Mayley-type cost of learning, are direct
analytic targets. Third, and most pointedly for the extended synthesis:
niche construction is the one channel that could enter the benefit term,
because what is transmitted then alters the environment itself, the sole
source of that term here. Whether a constructed niche can make socially
inherited information instructive, in exactly the sense measured in this
paper, is the natural next experiment.

\subsection{Limitations}

The landscape is a needle, chosen because it is the canonical case and because
it has no gradient to confound the measurement. Nothing here shows that the
result transfers to landscapes with structure, and the permissive reading may
well be an artefact of a fitness function whose only signal is exact match. The
target moves by single-locus flips at a constant rate; correlated or
autocorrelated change is not tested. Fitness is assessed once, at the end of
life, which as Best \cite{best} notes gives the learner the full benefit of
learning at no cost, so this model has no cost of learning in Mayley's
\cite{mayley} sense and cannot speak to genetic assimilation driven by such a
cost. The social signal is derived from successful individuals, which is a
strong selectivity assumption; and the thresholds of Table~\ref{tab:psistar}
are specific to $L = 20$, $G = 1000$ and one flipped locus per change event.
We have not varied $L$ or $G$. Equation~\eqref{eq:psistar} is a
one-generation, mean-field criterion evaluated at the population's mean
plasticity, which is why it misses the sign at the boundary. The fitness-informed
sample we use to stand in for Best's \cite{best} payoff-proportional selection
of models -- drawing uniformly from the fittest tenth rather than weighting by
fitness continuously -- is a tractable approximation to his procedure, not a literal replication of it.

Finally, the peripheral and central labels are ours, applied to $a$ and $b$ by
argument rather than by measurement. A referee is entitled to hold that a
conformity transform is not obviously more central than an attention gate, and
we would not resist the point. What the model establishes is that the two
parameters are separable and that they do different work; the mapping onto
cognitive architecture is an interpretation.

\section*{Data accessibility}
All simulation code, the archived outputs behind every number
reported, and the scripts that generate each figure are publicly available
at \url{https://github.com/drnamlabs-gif/forces-at-hand}; an archived copy
with a DOI will be deposited at Zenodo on acceptance. Archived outputs were
produced with Python 3.13 and NumPy 2.4.3 under fixed random seeds. Appendix~\ref{app:bestle} contains the demonstrator-sampling
comparison referred to in the text.

\section*{Funding}
This work received no specific funding. It originates in the
author's doctoral research and was completed independently; the author's
postdoctoral position is supported by the Army Research Laboratory under
contract no.\ W911NF-24-1-0041, which did not fund this study.

\section*{Ethics}
This study is entirely computational and involved no human
participants, animal subjects or fieldwork.

\section*{Acknowledgements}
The author thanks the University of Vermont, where this work was
carried out during the author's postdoctoral research.

\section*{Competing interests}
The author declares no competing interests.

\appendix
\setcounter{table}{0}\renewcommand{\thetable}{A\arabic{table}}
\setcounter{figure}{0}\renewcommand{\thefigure}{A\arabic{figure}}
\setcounter{equation}{0}\renewcommand{\theequation}{A\arabic{equation}}

\section{Can the social channel alone sustain adaptation? The demonstrator sample decides}
\label{app:bestle}

This appendix reports the experiment, referred to in the main text, that
resolves the disagreement between Best \cite{best} and Le \cite{le19} over
whether social learning alone can guide a population to its target. It uses the
second implementation of the model (Section~\ref{sec:results}): learning
budget $G=300$, $k$ demonstrators sampled per learner, fidelity $f=1$,
$b=0$, initial $p(?)=0.5$, $N=1000$, $L=20$, 200 generations, 12 independent
runs per condition, NumPy 2.4.3 with fixed random seeds.

Best's report \cite{best} that social learning alone finds the target, and
Le's report \cite{le19} that it does not, were never run against each other on the same code, with everything but the channel held fixed. We do that
here, inside equation~\eqref{eq:periph}, by fixing $a=1,\,b=0$ -- learners
guess purely from the social signal, with no individual contribution at all --
and varying only the sampling rule for $c_j$.

\begin{table}[htbp]\centering\small
\caption{Seven regimes on the static landscape, $L=20$, $G=300$, $N=1000$,
200 generations, 12 runs. \textsc{switch} is the original mechanism of Le \cite{le19},
discontinuous mechanism (oracle-gated copying), re-implemented as a separate
baseline rather than a point in equation~\eqref{eq:periph}.}
\label{tab:r5compare}
\resizebox{\textwidth}{!}{%
\begin{tabular}{@{}lrrrrr@{}}
\toprule
configuration & mean $F$ & final $p(?)$ & final $p(1)$ & solved / 12 & first solved (gen.)\\
\midrule
evolution alone (no plasticity) & $4.16$ & $0.000$ & $0.572$ & $2$ & $144$\\
individual learning alone, $a=0$ & $19.69$ & $0.094$ & $0.906$ & $12$ & $0$\\
\textbf{social learning alone, $k{=}1$, \textsc{parent}} & $\mathbf{1.00}$ & $0.515$ & $0.237$ & $\mathbf{0}$ & --\\
\textbf{social learning alone, $k{=}5$, fittest tenth} & $\mathbf{16.78}$ & $0.482$ & $0.470$ & $\mathbf{10}$ & $8$\\
blend, $a=0.8$, $k=5$, fittest tenth & $18.26$ & $0.455$ & $0.519$ & $12$ & $8$\\
blend, $a$ evolves, $k=5$, fittest tenth & $19.83$ & $0.414$ & $0.586$ & $12$ & $2$\\
\textsc{switch} (original mechanism of \cite{le19}) & $19.33$ & $0.530$ & $0.470$ & $12$ & $0$\\
\bottomrule
\end{tabular}}
\end{table}

\begin{figure}[htbp]\centering
\includegraphics[width=\textwidth]{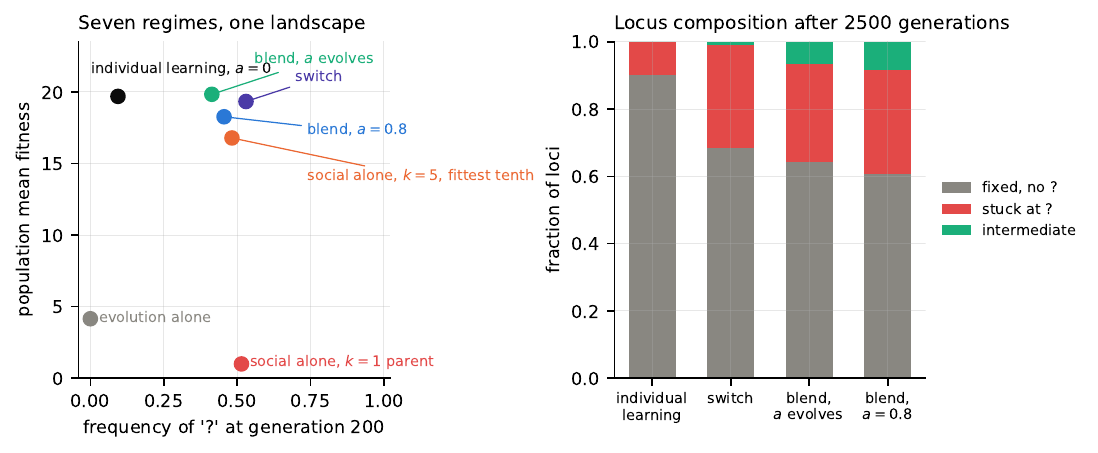}
\caption{The seven regimes of Table~\ref{tab:r5compare} on the
(plasticity, fitness) plane, and the locus composition each reaches after
2500 generations (Table~\ref{tab:r5long}).}
\label{fig:compare}
\end{figure}

The middle two rows differ only in $k$ and the sampling rule, with $a=1$,
$b=0$, fidelity, cost and everything else held identical. $k=1$ under
\textsc{parent} reproduces Le's result: not one of twelve runs finds the
target. $k=5$ drawn from the fittest tenth reproduces Best's: ten of twelve
runs find it, with a median first solution at generation eight. The two reports that stood as a contradiction for twenty years are two
points on one axis, \textbf{whether the demonstrator sample carries
variance}. A single inherited demonstrator is a conduit: it transmits a value
but manufactures none, so a population without individual guessing has no
source of new information and cannot discover anything it has not already
discovered. A sample of five drawn from a fitness-selected pool is a search
process in its own right, with variance across generations in who belongs to
the pool; that variance supplies exactly the exploration that a single
demonstrator cannot, which is sufficient, on this landscape, to find the
target with no individual guessing at all. The statement ``social learning
cannot learn anything that has not already been learned'' is therefore true
of the $k=1$ row and false of the $k=5$ row, and the field's disagreement was
over which row its intuitions were about.

Table~\ref{tab:r5long} extends all four plastic regimes to 2500 generations.

\begin{table}[htbp]\centering\small
\caption{After 2500 generations. ``Stuck'' is the fraction of loci still
undetermined; ``free'' is the fraction fixed to a determined value;
``mid'' is the fraction at an intermediate frequency across the population.}
\label{tab:r5long}
\begin{tabular}{@{}lrrrr@{}}
\toprule
configuration & population $p(?)$ & free & \textbf{stuck} & intermediate\\
\midrule
individual learning alone & $0.100$ & $0.900$ & $\mathbf{0.100}$ & $0.000$\\
\textsc{switch} \cite{le19} & $0.310$ & $0.683$ & $\mathbf{0.308}$ & $0.008$\\
blend, $a$ evolves & $0.327$ & $0.642$ & $\mathbf{0.292}$ & $0.067$\\
blend, $a=0.8$ & $0.354$ & $0.608$ & $\mathbf{0.308}$ & $0.083$\\
\bottomrule
\end{tabular}
\end{table}

Every regime with a social channel converges to roughly three times the
asocial stuck-locus fraction, whichever mechanism supplies the channel, and
the population-level frequency of ``?'' tracks the stuck fraction almost
exactly in every row. Section~\ref{sec:r7} argues this equality is itself
diagnostic.

This has a direct consequence for how earlier results, including Le's \cite{le19}, should be read. A run that ends with half its loci undetermined,
having been initialised with half its loci undetermined, is not evidence that
genes have stopped assimilating because plasticity became valuable. It is
evidence that selection stopped discriminating. The distinction is invisible to
the frequency diagnostic and immediate under $\beta_q$.


\begin{thebibliography}{99}
\bibitem{westeberhard03} M. J. West-Eberhard (2003). \emph{Developmental
  Plasticity and Evolution}. New York: Oxford University Press.
\bibitem{laland14} K. Laland, T. Uller, M. Feldman, K. Sterelny,
  G. B. M\"uller, A. Moczek, E. Jablonka, J. Odling-Smee, G. A. Wray,
  H. E. Hoekstra, D. J. Futuyma, R. E. Lenski, T. F. C. Mackay, D. Schluter and
  J. E. Strassmann (2014). Does evolutionary theory need a rethink?
  \emph{Nature} \textbf{514}, 161--164. (doi:10.1038/514161a)
\bibitem{laland15} K. N. Laland, T. Uller, M. W. Feldman, K. Sterelny,
  G. B. M\"uller, A. Moczek, E. Jablonka and J. Odling-Smee (2015). The
  extended evolutionary synthesis: its structure, assumptions and predictions.
  \emph{Proceedings of the Royal Society B} \textbf{282}, 20151019. (doi:10.1098/rspb.2015.1019)
\bibitem{pfennig21} D. W. Pfennig (ed.) (2021). \emph{Phenotypic Plasticity
  \& Evolution: Causes, Consequences, Controversies}. Boca Raton: CRC Press.
\bibitem{levis16} N. A. Levis and D. W. Pfennig (2016). Evaluating
  `plasticity-first' evolution in nature: key criteria and empirical
  approaches. \emph{Trends in Ecology \& Evolution} \textbf{31}(7), 563--574.
\bibitem{levis20} N. A. Levis and D. W. Pfennig (2020). Plasticity-led
  evolution: a survey of developmental mechanisms and empirical tests.
  \emph{Evolution \& Development} \textbf{22}(1--2), 71--87. (doi:10.1111/ede.12309)
\bibitem{uller19} T. Uller and K. N. Laland (eds) (2019). \emph{Evolutionary
  Causation: Biological and Philosophical Reflections}. Cambridge, MA: MIT
  Press.
\bibitem{calcott} B. Calcott (2017). Causal specificity and the
  instructive--permissive distinction. \emph{Biology \& Philosophy}
  \textbf{32}(4), 481--505.
\bibitem{bourrat} P. Bourrat (2019). On Calcott's permissive and
  instructive cause distinction. \emph{Biology \& Philosophy}
  \textbf{34}(1), 1--12.
\bibitem{gilbert} S. F. Gilbert (2013). \emph{Developmental Biology},
  10th edn. Sunderland, MA: Sinauer Associates.
\bibitem{baldwin} J. M. Baldwin (1896). A new factor in evolution.
  \emph{American Naturalist} \textbf{30}, 441--451.
\bibitem{crispo07} E. Crispo (2007). The Baldwin effect and genetic
  assimilation: revisiting two mechanisms of evolutionary change mediated by
  phenotypic plasticity. \emph{Evolution} \textbf{61}(11), 2469--2479. (doi:10.1111/j.1558-5646.2007.00203.x)
\bibitem{scheiner17} S. M. Scheiner, M. Barfield and R. D. Holt (2017). The
  genetics of phenotypic plasticity. XV. Genetic assimilation, the Baldwin
  effect, and evolutionary rescue. \emph{Ecology and Evolution} \textbf{7},
  8788--8803. (doi:10.1002/ece3.3429)
\bibitem{danchin11} \'E. Danchin, A. Charmantier, F. A. Champagne,
  A. Mesoudi, B. Pujol and S. Blanchet (2011). Beyond DNA: integrating
  inclusive inheritance into an extended theory of evolution. \emph{Nature
  Reviews Genetics} \textbf{12}(7), 475--486. (doi:10.1038/nrg3028)
\bibitem{hn} G. E. Hinton and S. J. Nowlan (1987). How learning can guide
  evolution. \emph{Complex Systems} \textbf{1}, 495--502.
\bibitem{nolfi94} S. Nolfi, J. L. Elman and D. Parisi (1994). Learning and
  evolution in neural networks. \emph{Adaptive Behavior} \textbf{3}(1), 5--28.
\bibitem{belew} R. K. Belew (1990). Evolution, learning and culture:
  computational metaphors for adaptive algorithms.
  \emph{Complex Systems} \textbf{4}, 11--49.
\bibitem{best} M. L. Best (1999). How culture can guide evolution: an inquiry
  into gene/meme enhancement and opposition. \emph{Adaptive Behavior}
  \textbf{7}, 289--306.
\bibitem{le19} N. Le (2019). Organic selection and social heredity: the
  original Baldwin effect revisited. In \emph{Proc. Artificial Life Conference
  (ALIFE 2019)}, 515--522. MIT Press.
\bibitem{le19gecco} N. Le, M. O'Neill and A. Brabazon (2019). Evolutionary
  consequences of learning strategies in a dynamic rugged landscape. In
  \emph{Proc. Genetic and Evolutionary Computation Conference (GECCO '19)},
  812--819. ACM. (doi:10.1145/3321707.3321741)
\bibitem{le20evoapps} N. Le, A. Brabazon and M. O'Neill (2020). Social
  learning vs self-teaching in a multi-agent neural network system. In
  \emph{Applications of Evolutionary Computation (EvoApplications 2020)},
  Lecture Notes in Computer Science vol.\ 12104, 354--368. Springer. (doi:10.1007/978-3-030-43722-0\_23)
\bibitem{heinerman} J. Heinerman, M. Drupsteen and A. E. Eiben (2015). Three-fold
  adaptivity in groups of robots: the effect of social learning. In
  \emph{Proc. Genetic and Evolutionary Computation Conference (GECCO
  2015)}, pp.\ 177--183. ACM. (doi:10.1145/2739480.2754743)
\bibitem{heinerman2} J. Heinerman, M. Rango and A. E. Eiben (2015). Evolution, individual
  learning, and social learning in a swarm of real robots. In \emph{Proc.
  IEEE Symposium Series on Computational Intelligence (SSCI 2015)},
  pp.\ 1055--1062. IEEE. (doi:10.1109/SSCI.2015.152)
\bibitem{bredeche} N. Bredeche and N. Fontbonne (2022). Social learning in
  swarm robotics. \emph{Philosophical Transactions of the Royal Society B}
  \textbf{377}, 20200309. (doi:10.1098/rstb.2020.0309)
\bibitem{heyes12} C. Heyes (2012). What's social about social learning?
  \emph{Journal of Comparative Psychology} \textbf{126}(2), 193--202.
\bibitem{futuyma17} D. J. Futuyma (2017). Evolutionary biology today and the
  call for an extended synthesis. \emph{Interface Focus} \textbf{7}, 20160145. (doi:10.1098/rsfs.2016.0145)
\bibitem{woodward10} J. Woodward (2010). Causation in biology: stability,
  specificity, and the choice of levels of explanation. \emph{Biology \&
  Philosophy} \textbf{25}, 287--318.
\bibitem{boyd85} R. Boyd and P. J. Richerson (1985). \emph{Culture and the
  Evolutionary Process}. Chicago: University of Chicago Press.
\bibitem{rogers} A. R. Rogers (1988). Does biology constrain culture?
  \emph{American Anthropologist} \textbf{90}(4), 819--831.
\bibitem{borenstein08} E. Borenstein, M. W. Feldman and K. Aoki (2008).
  Evolution of learning in fluctuating environments: when selection favors
  both social and exploratory individual learning. \emph{Evolution}
  \textbf{62}(3), 586--602. (doi:10.1111/j.1558-5646.2007.00313.x)
\bibitem{ram18} Y. Ram, U. Liberman and M. W. Feldman (2018). Evolution of
  vertical and oblique transmission under fluctuating selection.
  \emph{Proceedings of the National Academy of Sciences USA} \textbf{115}(6),
  E1174--E1183. (doi:10.1073/pnas.1719171115)
\bibitem{aokifeldman14} K. Aoki and M. W. Feldman (2014). Evolution of
  learning strategies in temporally and spatially variable environments: a
  review of theory. \emph{Theoretical Population Biology} \textbf{91}, 3--19.
\bibitem{heyes} C. Heyes, N. Chater and D. M. Dwyer (2020). Sinking in: the peripheral
  Baldwinisation of human cognition. \emph{Trends in Cognitive Sciences}
  \textbf{24}(11), 884--899. (doi:10.1016/j.tics.2020.08.006)
\bibitem{olsson} A. Olsson, E. Knapska and B. Lindstr\"om (2020). The neural
  and computational systems of social learning. \emph{Nature Reviews
  Neuroscience} \textbf{21}(4), 197--212.
\bibitem{fontanari} J. F. Fontanari and R. Meir (1990). The effect of learning
  on the evolution of asexual populations. \emph{Complex Systems} \textbf{4},
  401--414.
\bibitem{harvey} I. Harvey (1993). The puzzle of the persistent question marks: a case
  study of genetic drift. In \emph{Genetic Algorithms: Proceedings of the
  Fifth International Conference (GA93)} (ed.\ S. Forrest). San Mateo, CA:
  Morgan Kaufmann.
\bibitem{ancel00} L. W. Ancel (2000). Undermining the Baldwin expediting
  effect: does phenotypic plasticity accelerate evolution?
  \emph{Theoretical Population Biology} \textbf{58}(4), 307--319.
\bibitem{mayley} G. Mayley (1996). Landscapes, learning costs and genetic
  assimilation. \emph{Evolutionary Computation} \textbf{4}, 213--234.
\bibitem{morgan} T. J. H. Morgan, J. W. Suchow and T. L. Griffiths
  (2020). What the Baldwin effect affects depends on the nature of
  plasticity. \emph{Cognition} \textbf{197}, 104165. (doi:10.1016/j.cognition.2019.104165)
\bibitem{santos} M. Santos, E. Szathm\'ary and J. F. Fontanari (2015).
  Phenotypic plasticity, the Baldwin effect, and the speeding up of evolution:
  the computational roots of an illusion.
  \emph{Journal of Theoretical Biology} \textbf{371}, 127--136.
\bibitem{nolfi96} S. Nolfi and D. Parisi (1996). Learning to adapt to
  changing environments in evolving neural networks. \emph{Adaptive
  Behavior} \textbf{5}(1), 75--98.
\bibitem{hutchins} E. Hutchins and B. Hazlehurst (1992). Learning in the
  cultural process. In \emph{Artificial Life II} (eds C. G. Langton,
  C. Taylor, J. D. Farmer and S. Rasmussen), pp.\ 689--706. Redwood City,
  CA: Addison-Wesley.
\bibitem{denaro96} D. Denaro and D. Parisi (1996). Cultural evolution in a
  population of neural networks. In \emph{Neural Nets WIRN Vietri-96: Proc.
  8th Italian Workshop on Neural Nets}, Vietri sul Mare, Salerno, 100--111.
  Springer.
\bibitem{curran07} D. Curran and C. O'Riordan (2007). The effects of
  cultural learning in populations of neural networks. \emph{Artificial
  Life} \textbf{13}(1), 45--67. (doi:10.1162/artl.2007.13.1.45)
\bibitem{chisausky25} J. Chisausky, I. M. Daras, F. J. Weissing and
  M. Kozielska (2025). A neural network model for the evolution of
  reconstructive social learning. \emph{Scientific Reports} \textbf{15},
  14977. (doi:10.1038/s41598-025-97492-4)
\bibitem{cook24} J. Cook, C. Lu, E. Hughes, J. Z. Leibo and J. Foerster
  (2024). Artificial generational intelligence: cultural accumulation in
  reinforcement learning. In \emph{Advances in Neural Information Processing
  Systems 37 (NeurIPS 2024)}.
\bibitem{lande83} R. Lande and S. J. Arnold (1983). The measurement of
  selection on correlated characters. \emph{Evolution} \textbf{37}(6),
  1210--1226.
\bibitem{kingsolver01} J. G. Kingsolver, H. E. Hoekstra, J. M. Hoekstra,
  D. Berrigan, S. N. Vignieri, C. E. Hill, A. Hoang, P. Gibert and P. Beerli
  (2001). The strength of phenotypic selection in natural populations.
  \emph{American Naturalist} \textbf{157}(3), 245--261.
\bibitem{enquist07} M. Enquist, K. Eriksson and S. Ghirlanda (2007).
  Critical social learning: a solution to Rogers' paradox of nonadaptive
  culture. \emph{American Anthropologist} \textbf{109}(4), 727--734.
\bibitem{chevin10} L.-M. Chevin, R. Lande and G. M. Mace (2010).
  Adaptation, plasticity, and extinction in a changing environment: towards
  a predictive theory. \emph{PLoS Biology} \textbf{8}(4), e1000357.
  (doi:10.1371/journal.pbio.1000357)
\end{thebibliography}
\end{document}